\documentclass[aps, twocolumn,superscriptaddress]{revtex4-2} \usepackage{natbib}
\usepackage{import}
\usepackage{xifthen}
\usepackage{transparent}
\usepackage[colorlinks=true]{hyperref}
\usepackage{array}

\usepackage{verbatim, bm}
\usepackage{physics}
\usepackage{textcomp}
\usepackage{nicefrac}
\usepackage{mathrsfs}  
\usepackage[normalem]{ulem}
\usepackage{mathtools}
\usepackage{amsfonts}
\usepackage{amssymb}
\usepackage{color}
\usepackage{amsthm}
\usepackage{slashed}
\usepackage{amsmath}
\usepackage{wrapfig}
\usepackage{dsfont}
\usepackage{esint}

\hypersetup{
    unicode=false,          % non-Latin characters 
    pdftoolbar=true,        % show Acrobat
    pdfmenubar=true,        % show Acrobat 
    pdffitwindow=false,     % window fit to page when opened
    pdfstartview={FitH},    % fits the width of the page to the window
    pdfsubject={},          % subject of the document
    pdfcreator={},          % creator of the document
    pdfproducer={},         % producer of the document
    pdfkeywords={} {} {},   % list of keywords
    pdfnewwindow=true,      % links in new window
    colorlinks=true,        % false: boxed links; true: colored links
    linkcolor=magenta,      %red,          % color of internal links (change box color with linkbordercolor)
    citecolor=blue,         % color of links to bibliography
    filecolor=magenta,      % color of file links
    urlcolor=blue           % color of external links
} 
\allowdisplaybreaks

\renewcommand{\tilde}{\widetilde}
\renewcommand{\hat}{\widehat}

\newcommand{\pdagger}{{\phantom{\dagger}}}

\newcommand{\UU}{\operatorname{U}}

\newcommand{\SU}{\operatorname{SU}}

\begin{document}
\title{Emergent polaronic correlations in doped spin liquids}
\author{Leyna Shackleton}
\affiliation{Department of Physics, Massachusetts Institute of Technology, Cambridge MA 02139, USA}
\author{Shiwei Zhang}
\affiliation{Center for Computational Quantum Physics, Flatiron Institute, New York, NY, 10010, USA}
\date{\today\vspace{0.4in}}

\begin{abstract}
  The interplay between spin and charge degrees of freedom arising from doping a Mott insulating quantum spin liquid (QSL) has been a topic of research for several decades. Calculating properties of these fractionalized metallic states in single-band models are generally restricted to mean-field patron descriptions and small fluctuations around these states, which are insufficient for quantitative comparison of observables to measurements performed in strongly-correlated systems. In this work, we numerically study a class of correlated electronic wavefunctions which support fractionalized spin and charge excitations and which fully take into account gauge fluctuations through the enforcement of local Hilbert space constraints. By optimizing the energy of these wavefunctions against the hole-doped Fermi Hubbard Hamiltonian, %model, 
  we obtain a variational ansatz for describing the low-energy physics of this model. We compare measurements of hole-induced spin-spin correlation functions to measurements taken in low temperature cold-atom simulations of the Hubbard model and find quantitative agreement between the two. %- in 
  In particular, we demonstrate the emergence of magnetic polaron correlations in these metallic states.
\end{abstract}

\maketitle
\section{Introduction}
Quantum spin liquids (QSLs) are highly entangled ground states of insulating quantum antiferromagnets, which exhibit no long-range order and instead host fractionalized spin-$1 / 2$ spinon excitations~\cite{savary2016, broholm2020}. In elucidating properties of these exotic phases, parton constructions have played a crucial role - by writing the quantum spin operator as a pair of spinon operators, one can reformulate the Hamiltonians of these frustrated spin models as theories of interacting spinons coupled to emergent gauge fields. Mean-field theories of these spinons give useful starting points for describing and classifying QSL phases~\cite{wen2002}.  In these constructions, an important feature is the ability to non-perturbatively go beyond mean-field and fully account for gauge fluctuations on top of a non-interacting state through a Gutzwiller projection, which can be implemented numerically with Monte Carlo sampling. The ability to construct such correlated variational wavefunctions has played an essential part in determining energetics in frustrated Heisenberg models as well as quantitatively comparing observables to more unbiased studies.

Moving away from the Heisenberg limit, a natural question is how to incorporate charge fluctuations in a fractionalized spin liquid state, either by considering a Mott insulator with a non-infinite charge gap or a doped metallic spin liquid that retains deconfined fractionalized excitations. QSLs with finite charge gaps are of %or 
relevance to fractionalized phases that appear at half-filling near metal-insulator transitions, which are believed to exist in triangular lattice Hubbard models~\cite{szasz2020} as well as triangular Hubbard-Hofstadter models~\cite{kuhlenkamp2024}. In regards to the latter scenario of fractionalized metallic states, the multitude of strongly correlated phenomena that arise as a result of doping interaction-driven Mott insulators have long been a source of intense theoretical and experimental study, with a particular emphasis on the phenomenology of the high-temperature cuprate superconductors. Parton theories of electron fractionalization~\cite{lee2006}, where the electron operator is represented as a product of a charge-neutral spinon and spinless chargon, naturally admit a connection to an insulating QSL phase. However, these theories are often studied at a mean-field level, with interactions included perturbatively. Prior variational studies of these doped spin liquids with Gutzwiller projections~\cite{ivanov2000, ivanov2003} require a finite holon condensate, which leads to confinement and symmetry-broken phases. The ability to non-perturbatively study the effects of gauge fluctuations on the deconfined phases of doped spin liquids has remained an open problem for several decades. 

%The inability to numerically study deconfined variational states is an obstacle in constructing energetically-competitive variational wavefunctions for spin liquids that emerge near metal-insulator transitions, as well as understanding non-perturbative effects of gauge fluctuations on Fermi liquids with fractionalized excitations.

In this work, we address this problem by analyzing a class of variational wavefunctions that fully take into account gauge fluctuations and produce correlated electronic wavefunctions. As a starting point, we use recently-developed parton constructions using ancilla qubits~\cite{zhang2020} for capturing fractionalized spin and charge excitations in single-band models. In contrast to more conventional spinon/holon mean field theories, these constructions admit fractionalized Fermi liquid (FL*) phases with coherent electron-like excitations at a mean-field level. Moreover, this construction admits fully-fermionic mean-field ansatzes, which allows us to leverage the expressivity and computational efficiency of Slater determinant wavefunctions. With this, we can non-perturbatively account for gauge fluctuations through imposing local constraints, analogous to a Gutzwiller-projected spinon mean-field wavefunction used for describing insulating spin liquid states. This allows us to compute observables of these states.

Through analyzing these variational ansatzes, we make two notable observations. The first is that strong antiferromagnetic correlations, which for our choice of spin liquid ansatz are known to exist at half filling, persist upon hole doping. The second observation is the emergence of polaronic correlations at low doping, where a hole becomes dressed with a cloud of magnetic correlations that oppose the background antiferromagnetic fluctuations. Remarkably, we find that by variationally optimizing the energy of our fractionalized states against the square lattice Fermi-Hubbard model, these polaronic correlations agree quantitatively with recent cold-atom simulations of the Fermi-Hubbard model~\cite{koepsell2021}.

%. We provide a quantitative analysis of the applicability of these FL* wavefunctions in capturing low-temperature properties of the Fermi-Hubbard model. In particular, we demonstrate the emergence of polaronic correlations at small hole doping through the evaluation of local multi-point correlation functions, which agrees quantitatively with cold-atom simulations of the Fermi-Hubbard model~\cite{koepsell2021}.

We note that auxiliary degrees of freedom, similar to the ones used in our parton construction, have been used to great success in enhancing the expressivity of variational wavefunctions. This includes recent applications of auxiliary field quantum Monte Carlo (AFQMC) to parameterize variational wavefunctions~\cite{sorella2022, levy2024}, neural network hidden fermion determinental states~\cite{robledomoreno2022, gauvin-ndiaye2023}, fermionic shadow wavefunctions~\cite{calcavecchia2015}, as well as ghost Gutzwiller approximations~\cite{lanata2017}. While our motivation differs from these works - our interest is in describing a particular form of fractionalization which is best captured with auxiliary degrees of freedom, rather than bolstering the expressivity and reducing bias in variational wavefunctions - the possibility of developing explicit order detection through the use of auxiliary degrees of freedom is a promising future direction of research.
\section{Description of Variational Wavefunctions} 
In this section, we describe the class of wavefunctions under consideration in this work. As discussed previously, the goal of these wavefunctions is to capture charge fluctuations - induced by either %by 
doping or finite Hubbard repulsion - on top of a Mott insulating spin liquid state, in a manner which retains fractionalized spinon excitations. We accomplish this through an ancilla construction, first proposed in~\cite{zhang2020a}. While this pattern of fractionalization is more complicated than a more standard rewriting of an electron in terms of a charge-neutral spinon and a spinless holon, this approach has a number of advantages.
%which we elaborate on here. 
First, it retains electron-like excitations at a mean-field level, and hence a mean-field prediction for the electron Green's function can be obtained directly. These explicit electron-like degrees of freedom also admit a simple procedure for doping the state, as one can explicitly dope electrons and holes into the system. Secondly, because the parton construction exclusively uses fermionic degrees of freedom, the mean-field ansatz can effectively be parameterized using Slater determinant states, which turns out to be essential in numerically going beyond mean-field.

%It is useful to develop an analogy between our wavefunctions and Gutzwiller-projected spinon wavefunctions. In capturing the behavior of a spin liquid in a quantum antiferromagnet, one can start with a mean-field description in terms of fermionic spinons at half-filling $\ket{\psi_{\text{MF}}}$, which must be coupled to a dynamical gauge field in order to faithfully represent the physical Hilbert space. Integrating out this gauge field can be equivalently accomplished by performing a Gutzwiller projection $\ket{\psi} = \mathcal{P}_G \ket{\psi_{\text{MF}}}$ which projects out all doubly-occupied states. Formally this only accounts for the temporal fluctuations of the gauge field, as the spatial components remain fixed. Including charge fluctuations on top of this, either by doping the system or allowing for doublon/holon fluctuations at half-filling, is non-trivial. The problem is less challenging if one assumes condensation of a spin-neutral charged holon, in which case the electron and spinon become equivalent and one can describe these states by either doping $\ket{\psi_{\text{MF}}}$ or by softening the strength of the Gutzwiller projection. This will generically lead to symmetry-broken states such as $d$-wave superconductivity, which may be desirable if one intends to study these phases. Our goal, however, is to develop a wavefunction that retains spin/charge fractionalization and capture phases which support deconfined gauge fluctuations.

It is instructive to consider how one might construct correlated wavefunctions using conventional spinon/holon fractionalization. We rewrite the electron in terms of a fermionic spinon and a bosonic holon
\begin{equation}
    \begin{aligned}
        c^\dagger_{i\sigma} = f^\dagger_{i \sigma} b_i\,,
    \end{aligned}
\end{equation}
which enlarges our Hilbert space and introduces a $\UU(1)$ gauge redundancy. The physical Hilbert space is defined by the constraint
\begin{equation}
    \begin{aligned}
        f_{i\sigma}^\dagger f_{i\sigma}  + b_i^\dagger b_i = 1\,. 
        \label{eq:spinonHolonConstraint}
    \end{aligned}
\end{equation}
By constructing a mean-field wavefunction $\ket{\psi_{\text{MF}}} \equiv \ket{\psi^f_{\text{MF}}} \otimes \ket{\psi^b_{\text{MF}}}$ and sampling states $\ket{n^f_i, n^b_i}$ on the eigenbasis of $\left(f_{i\sigma}^\dagger f_{i\sigma}\,, b_i^\dagger b_i\right)$ using Monte Carlo methods, one can exactly enforce Eq.~\ref{eq:spinonHolonConstraint} exactly and obtain a correlated electronic wavefunction. Although this is a well-defined procedure, an obstacle in this approach is constructing an appropriate trial mean-field ansatz for the holons $b_i$ which retains deconfined excitations while supporting finite doping. A more severe issue is that enforcing the permutation symmetry of noninteracting bosonic wavefunctions naturally leads to the computation of matrix permanents when computing wavefunction overlaps, which is exponentially most costly to compute than the determinants of fermionic wavefunctions. As a result, the evaluation of the correlated wavefunction becomes intractable numerically. Hence, a formulation that uses exclusively fermionic degrees of freedom on a mean-field level is more likely to admit interpretable and practical wavefunctions through the use of Slater determinant states.

\begin{figure*}
    \centering
    \includegraphics[width=\textwidth]{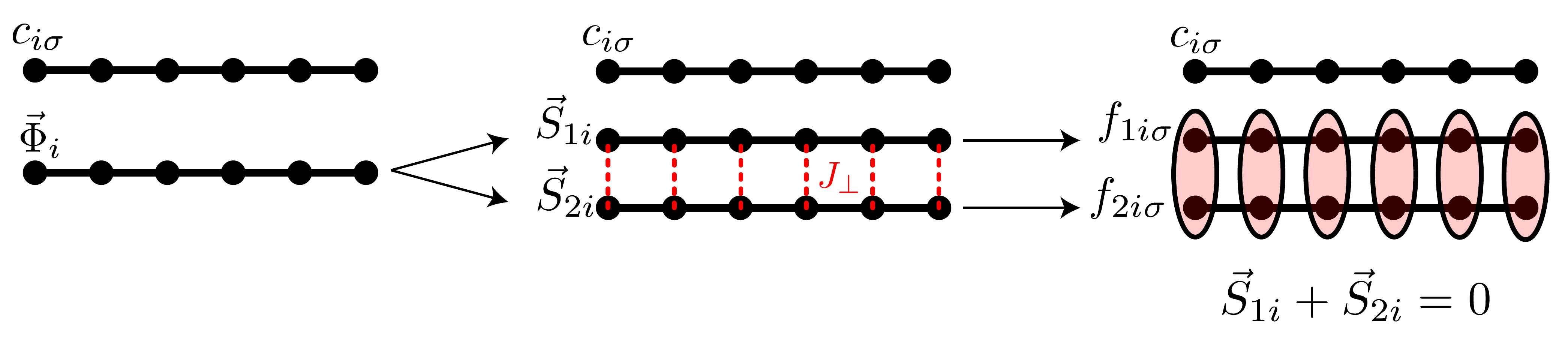}
    \caption{%We present a summary 
    Summary of the parton construction used in our variational wavefunctions. A spin-$1$ paramagnon $\vec{\Phi}$ (left) %(right) 
    is re-expressed in terms of two antiferromagnetically-coupled spin-$1/2$ moments (center). These spins are then fractionalized into Abrikosov fermions (right) %(left) 
    which are allowed to hybridize with the physical electrons. The additional degrees of freedom are obtained by projecting into the paramagnon ground state, which consists of spin singlets on each site.}
    \label{fig:fractionalization}
\end{figure*}

Our alternative procedure involves fractionalization of a charge-neutral spin-1 paramagnon, as described in~\cite{zhang2020, zhang2020a, mascot2022} and illustrated in Fig.~\ref{fig:fractionalization}. We refer the reader to~\cite{mascot2022} for an explicit derivation of the wavefunction starting from a repulsive Hubbard model and give a brief summary here. The singlet and triplet excitations of the spin-1 paramagnon are captured by two layers of antiferromagnetically coupled spin-1 / 2 moments, which are then fractionalized into fermionic spinons. The resulting degrees of freedom involve three sets of spinful fermions: the physical electrons $c$, and two layers of ``ancilla'' fermions $f_1$ and $f_2$. For our particular mean-field ansatz, which describes a fractionalized Fermi liquid (FL*), the $f_2$ fermions are decoupled and form a spin liquid, whereas the $c$ and $f_1$ fermions hybridize and form electron-like excitations. 
%The formation of such an FL* state can be motivated energetically by observing that the coupling between the spin-1 paramagnon and the spin of the electron naturally induces antiferromagnetic (ferromagnetic) Kondo couplings on the first (second) layer of ancilla spins, promoting formation of a heavy Fermi liquid-like state between the $c$ and $f_1$ layers with a decoupled second layer of spins.
Our correlated wavefunctions in the physical Hilbert space of $c$ fermions is obtained by projecting the two ancilla layers into local rung singlets,
\begin{equation}
  \begin{aligned}
    \ket{\psi} \equiv \mathcal{P}_{SS} \left[ \ket{\psi^{\text{MF}}_{c, f_1}} \otimes \ket{\psi^{\text{MF}}_{f_2}} \right] \,,
    \label{eq:variationalWF}
  \end{aligned}
\end{equation}
and then integrating out the ancilla degrees of freedom.
%Intuitively, this rung singlet projection ``teleports'' or encodes the spin correlations of the spin liquid into the physical electron-like excitations~\cite{zhou2023}.

We will be interested in studying the the quantitative properties of these variational wavefunctions on the square lattice. Concretely, $\ket{\psi^{\text{MF}}_{c, f_1}}$ is the Slater determinant ground state of the Hamiltonian
\begin{equation}
\label{eq:H-cf1}
  \begin{aligned}
    &H_{c, f_1} \equiv - t_c \sum_{\langle i j \rangle} c_{i \sigma}^\dagger c_{j \sigma} - t_1 \sum_{\langle i j \rangle} f_{1, i \sigma}^\dagger f_{1, j \sigma}
    \\
    &+ \Phi \sum_i \left(c_{i \sigma}^\dagger f_{1, i \sigma}+ f_{1, i \sigma}^\dagger c_{i \sigma} \right) - \mu_f \sum_i f_{1, i \sigma}^\dagger f_{1, i \sigma} \,.
  \end{aligned}
\end{equation}
For modeling a Hamiltonian with $N_e$ electrons and $N$ sites, the total Hamiltonian is fixed to have $N + N_e$ and the chemical potential $\mu_f$ is tuned so that the average density in the $f_1$ layer is one electron per site. 

The Hamiltonian for the second ancilla layer is a $\pi$-flux fermionic spinon Hamiltonian at half-filling,
\begin{equation}
  \begin{aligned}
    &H_{f_2} = \sum_{ij} t_{ij} f_{2, i \sigma}^\dagger f_{2, j \sigma}\,,
    \\
    &t_{i, i + \hat{x}} = - t_{i + \hat{x}, i} = i\,,
    \\
    &t_{i, i + \hat{y}} = - t_{i + \hat{y}, i} = (-1)^{i_x} i\,.
  \end{aligned}
\end{equation}
Although such a spin liquid is believed to ultimately be unstable at low energy, recent symmetry analyses of these FL* phases with at $\pi$-flux spin liquid~\cite{christos2023} demonstrate instabilities N\'eel antiferromagnetism, charge density wave order, and $d$-wave superconductivity. This rich interplay between competing ordered phases motivates us to consider this ansatz.

The overall energy scale of $H_{f_2}$ is irrelevant to the final wavefunction. Moreover, we find that the parameters $t_c\,, t_1$ are not independent from each other; the dependence between hoppings on the $c$ layer and on the $f_1$ layer holds for generic-length hoppings and can be verified by numerical evaluation of the quantum geometric tensor. Taking into account the arbitrariness of the overall energy scale of $H_{c, f_1}$, there is only one independent parameter, which we take to be $\Phi$ and fix $t_c = -t_1 = 1$.
\section{Numerical Implementation}
Here, we briefly describe %the details of numerically evaluating 
how observables are evaluated 
with respect to the variational %Hamiltonian described 
wave function
in Eq.~\ref{eq:variationalWF} and defer a more thorough presentation to Appendix~\ref{app:numerics}. The projection operator $\mathcal{P}_{SS}$ acts as the identity on the physical Hilbert space, and within the ancilla subspace acts as
\begin{equation}
  \begin{aligned}
    \mathcal{P}_{SS} \equiv \otimes_i \left[ \ket{\uparrow \downarrow}_i - \ket{\downarrow \uparrow}_i \right] \left[ \bra{\uparrow \downarrow}_i - \bra{\downarrow \uparrow}_i \right] = \sum_{x, y} \ket{x}\bra{y}
    \label{eq:projection}
  \end{aligned}
\end{equation}
where $\ket{\uparrow \downarrow}_i \equiv f_{1, i\uparrow} f_{2, i\downarrow}\ket{0}$ and similarly for $\ket{\downarrow \uparrow}_i$. The states $\ket{x}\,, \ket{y}$ enumerate all real-space configurations of the ancilla fermions (in the $S^z$ basis) such that each ancilla site has one fermion and each rung has total $S^z$ equal to zero. Note that in the final equality we have ignored the relative minus signs in the spin singlet configurations, as it can be absorbed into the definition of the many-body basis $\ket{a}$.

As we demonstrate in Appendix~\ref{app:numerics}, implementing the full rung singlet projection leads to a sign problem. As such, analyzing these wavefunction ansatzes requires more care. To this end, we define an alternate projection operator parameterized by a variable $\beta$,
\begin{equation}
  \begin{aligned}
    \mathcal{P}_\beta &\equiv e^{-\beta \sum_i \left( \vb{S}_{f_1, i} + \vb{S}_{f_2, i} \right)^2 } \mathcal{P}_{S^z = 0}\,,
     \\
     \mathcal{P}_{S^z = 0} &\equiv \sum_{x} \ket{x}\bra{x}\,.
  \end{aligned}
\end{equation}
The operator $\mathcal{P}_{S^z=0}$ projects to a subspace where the total $S^z$ on each rung is zero, and can be implemented without a sign problem using standard variational Monte Carlo methods. To go beyond this and project out the residual rung triplet contributions $\ket{\uparrow \downarrow} + \ket{\downarrow \uparrow}$ will introduce a sign problem in our Monte Carlo sampling, the severity of which can be tuned by our parameter $\beta$. The full rung singlet projection is obtained in the limit $\beta \rightarrow \infty$. In practice, we find that the severity of the sign problem scales roughly with the strength of charge fluctuations, with the average sign decreasing upon doping the system or by reducing the size of the charge gap at half-filling. An important observation is that, while our fractionalized wavefunctions of interest are recovered in the $\beta\rightarrow \infty$ limit, we retain well-defined variational states for arbitrary $\beta$; in other words, observables calculated at finite $\beta$ constitute true physical observables with respect to a particular variational state, rather than approximated observables. On a field-theoretic level, the rung singlet projection is implemented by the introduction of a dynamical $\frac{\SU(2) \otimes \SU(2) \otimes \SU(2)}{\mathbb{Z}_2}$ gauge field; two of the $\SU(2)$ gauge fields implement the no-double-occupancy constraint on each of the ancilla layers, and the third projects into the rung singlet sector. Our approximation can be thought of as fully accounting for fluctuations of a $\frac{\SU(2) \otimes \SU(2) \otimes \UU(1)}{\mathbb{Z}_2}$ subgroup which only constrains the total $S^z$ on each rung to be zero, and the residual gauge fluctuations are integrated out gradually with increasing $\beta$. On the level of Hilbert spaces, however, we note a qualitative difference for finite $\beta$, as the corresponding projection operator does not fully disentangle the ancilla degrees of freedom from the physical ones. As a result, our variational state for finite $\beta$ should be more accurately thought of as a mixed state, $\rho \equiv \Tr_{f_1\,, f_2} \left[ \ket{\psi} \bra{\psi} \right]$, which becomes pure in the $\beta \rightarrow \infty$ limit.

\section{Results}

%\section{Energetics at half-filling}
Having demonstrated that these fractionalized states are accessible computationally, we now compute various properties of these states. While the parameters of these states could be set arbitrarily, we instead choose to fix the parameters by variationally optimizing them against the square lattice Fermi-Hubbard model. Doing so serves a dual purpose. Firstly, as a paradigmatic model of strongly-correlated electrons which exhibit a complex interplay between spin and charge degrees of freedom, we conjecture that choosing parameters that put our wavefunction ansatzes in the low-energy subspace of the Hubbard model will generate states with similarly rich correlations. Our second motivation is to investigate the ability of these variational states to capture low-temperature properties of the doped Hubbard model. Recent cold-atom simulations of the doped Hubbard model~\cite{koepsell2021} have demonstrated the emergence of polaronic correlations at low temperature, where holes at low doping become dressed with a cloud of magnetic correlations that oppose the background antiferromagnetic correlations. These polaronic features are suppressed with increasing doping, ultimately leading to a crossover to Fermi liquid-like correlations at high doping. Moreover, comparison of these multi-point correlation functions with a large class of wavefunction ansatzes shows that no single ansatz is capable of capturing the crossover from polaronic correlations at low doping to Fermi liquid-like at higher doping. By optimizing our variational states with respect to the Fermi-Hubbard model, we test whether these wavefunction anstazes are able to succeed in describing these polaronic correlations. 

We 
variationally %variational
optimize the energy of these FL* wavefunctions against the square lattice Fermi-Hubbard model,
\begin{equation}
    H = -t\sum_{\langle i j \rangle}c_{i\sigma}^\dagger c^\pdagger_{i\sigma} + U \sum_i n_{i\uparrow} n_{i\downarrow}\,.
\end{equation}
All simulations are performed on an $8 \times 8$ lattice with twist average boundary conditions~\cite{lin2001}, averaged over $356$ boundary conditions. Note that for finite-size systems, it is not always possible for an arbitrary boundary condition to tune the chemical potential $\mu_f$ in Eq.~\ref{eq:H-cf1} such that the density of $f_1$ electrons is fixed to $1$, as the filling may discontinuously jump from $n_{f_1} < 1$ to $n_{f_1} > 1$ through a level crossing. In these situations, we re-randomize boundary conditions until the filling constraint can be satisfied.

\subsection{Energetics at half-filling}

We briefly review the energetic performance of these wavefunctions at half-filling over a range of $U/t$, as the absence of a sign problem in conventional QMC simulations admits comparison of the variational energy with accurate and unbiased QMC estimates~\cite{qin2016}. Moreover, the weakness of our sign at half-filling allows for us to take the $\beta \rightarrow \infty$ limit and simulate our variational wavefunctions exactly. For all $U / t > 0$, the true ground state of the Hubbard model is known to be an antiferromagnetic insulator. As a result, a paramagnetic spin liquid ansatz will incur some energy cost; however, the power-law antiferromagnetic correlations of the $\pi$-flux spin liquid are known to produce a low-energy ansatz with a relative error of around 3\% in the Heisenberg limit~\cite{krivnov1994}. The purpose of this section is to analyze how our variational wavefunctions capture gapped charge fluctuations on top of this $\pi$-flux ansatz. 

For $U / t \rightarrow \infty$, the optimal wavefunction must have no double occupancy. This can be accomplished by sending $\Phi \rightarrow \infty$, where our variational wavefunction reduces to the standard Gutzwiller-projected spinon wavefunction. For finite $U / t$, we find remarkably that the ansatz $\Phi = U $ is nearly fully optimal over the full range of parameters considered, from $U / t = 2$ to $U / t = 12$. At a mean-field level, this seemingly implies our ansatz has a charge gap of magnitude $2U$. While this provides an attractive picture, similar to ghost Gutzwiller methods~\cite{lanata2017}, of an upper and lower Hubbard band on a mean-field level, the magnitude of the gap differs by a factor of 2 from a naive large-$U$ estimate of a charge gap $U$. This discrepancy can be resolved by remembering that the overall energy scale of our mean-field Hamiltonian is unfixed, and the fact that it is defined in a different Hilbert space than the physical Hamiltonian means that care must be taken when comparing the two. A careful perturbative analysis around the $U\,, \Phi=\infty$ limit performed in~\cite{zhou2023} demonstrates that the ansatz $\Phi=U$ for our mean-field choice correctly recovers the inverse Schrieffer-Wolff transformation away from the $\Phi=\infty$ limit. Alternatively, one can observe that the physical ground state energy and the mean-field ground state energy in the $U\,, \Phi=0$ limit only agree if one identifies $t_c = t_1 = t / 2$, which also resolves the numerical factor. The presence of electron hopping is known to substantially modify the charge gap in the Hubbard model from the zeroth-order prediction of $U$, with recent DQMC studies demonstrating that the charge gap is exponentially suppressed in $\sqrt{\frac{t}{U}}$ at small $U$ and crosses over to a more complex scaling predicted by the one-dimensional Bethe ansatz solution at large $U$~\cite{vitali2016}. While dynamical properties cannot easily be calculated within our framework, one may postulate that the gauge fluctuations work to renormalize our variational charge gap to one of the same magnitude as the true ground state. 

\begin{figure}[htpb]
  \centering
  \includegraphics[width=0.5\textwidth]{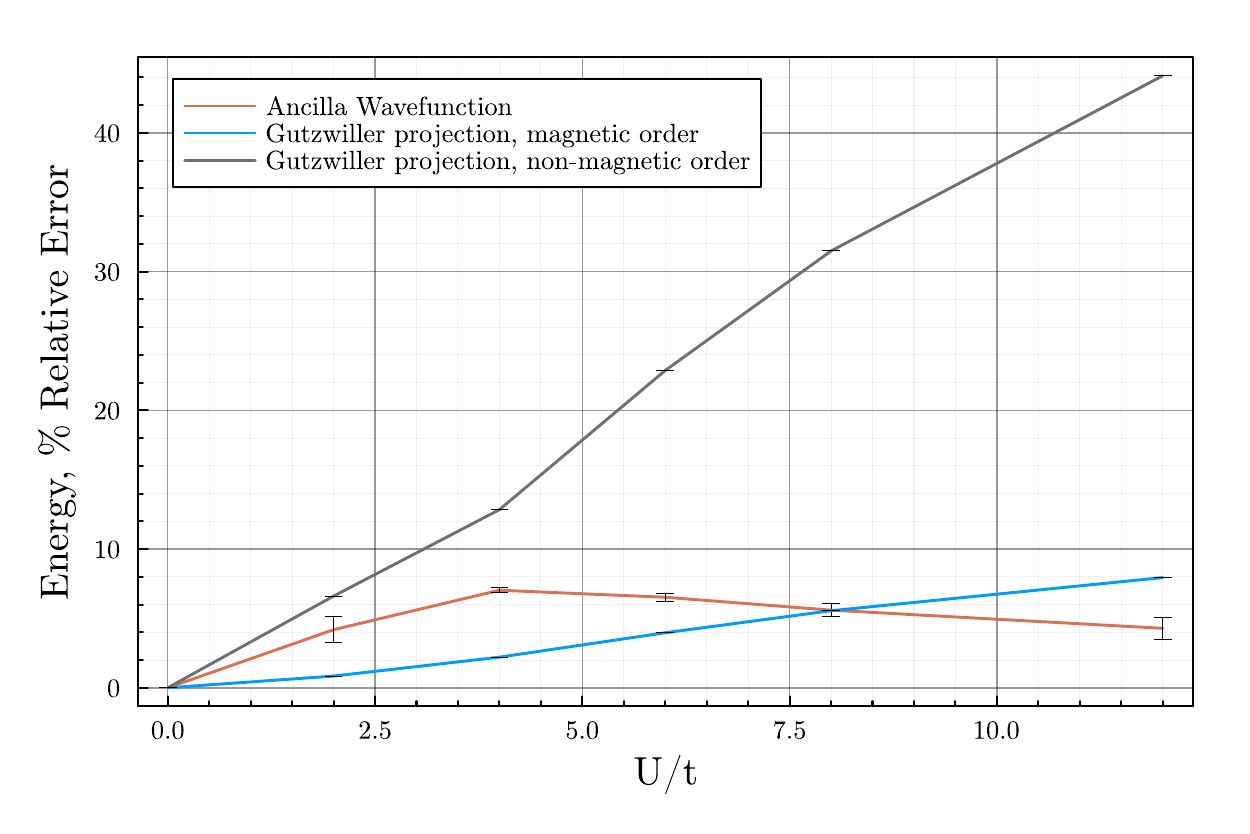}
  \caption{%We plot the relative 
  Relative error in the optimized energy of several few-parameter variational ansatzes for the Hubbard model at half-filling. The ancilla wavefunctions, which describe Mott-insulating $\pi$-flux states with a finite charge gap, are to a good accuracy optimized by setting the mean-field charge gap to the Hubbard repulsion $U$. Despite the simplicity of this ansatz, it performs well over the full range of $U/t$, giving significantly better results than a metallic Gutzwiller-projected free electron gas (with the strength of the projection as a free parameter) and comparable energetics to Gutzwiller-projected free electrons with a staggered magnetization (with the strength of the magnetization also optimized). }
  \label{fig:halfFilling}
\end{figure}

We emphasize that this essentially zero-parameter fit performs remarkably well, to around 5\% relative error to the true ground state over the full range of $U / t$. This ansatz performs worse for smaller $U / t$, which is consistent with the interpretation of the true ground state charge gap crossing over from Mott-like at large $U / t$ to a non-interacting bandgap driven by spin density wave order at small $U / t$. We plot the relative error to the true ground state as computed in AFQMC~\cite{qin2016} in Fig~\ref{fig:halfFilling}. We compare %contrast 
this with two %several 
other few-parameter variational wavefuncitons, %such as 
a Gutzwiller-projected paramagnetic Slater determinant (with one parameter determining the strength of the projection) and a Gutzwiller-projected Slater determinant with antiferromagnetic order in the $S^z$ direction (an additional parameter determines the strength of the order). 
The former is systematically less accurate. The latter gives slightly better energy at small and intermediate $U/t$, however it breaks translational symmetry, in contrast with our ansatz.
From this analysis, we conclude that these variational wavefunctions are able to reasonably capture gapped charge excitations on top of an insulating spin liquid state. 

%\section{Polaronic correlations away from half-filling}
\subsection{Polaronic correlations away from half-filling}

We now investigate the properties of local multi-point correlation functions in our variational wavefunctions away from half-filling. We choose parameters $U / t = 7.4$ and $U/t = 8.9$, with the goal of comparing to measurements in cold-atom systems~\cite{koepsell2021} which target %estimate 
this interaction strength. Following~\cite{koepsell2021}, we define a measure of the spin-spin correlations
\begin{equation}
\begin{aligned}
    C^c(\vb{r}_1, \vb{r_2}) &\equiv \langle \hat{S}^z_1 \hat{S}^z_2 \rangle / \left(\sigma\big(\hat{S}_2^z\big) \sigma\big(\hat{S}_1^z\big) \right)\,,
    \\
\sigma\big(\hat{S}_1^z\big) &\equiv \sqrt{\langle \hat{S}_1^{z^2} \rangle - \langle \hat{S}_1^z \rangle^2}\,.
\end{aligned}
\end{equation}
We also define a measure of hole-induced spin-spin correlations,
\begin{equation}
    \begin{aligned}
       & C^c_3(\vb{r}_1, \vb{r}_2, \vb{r}_3) = \langle \hat{h}_3 \hat{S}^z_2 \hat{S}^z_1 \rangle  %\,, removed. OK? 
       \\
       &-\langle \hat{h}_3 \rangle \langle \hat{S}^z_2 \hat{S}^z_1 \rangle - \langle\hat{S}^z_2\rangle \langle \hat{h}_3  \hat{S}^z_1 \rangle %\,,
       \\
       &-\langle\hat{S}^z_1\rangle \langle \hat{h}_3  \hat{S}^z_2 \rangle + 2  \langle \hat{h}_3 \rangle  \langle \hat{S}^z_2 \rangle \langle \hat{S}^z_1 \rangle  \,,
       \\
       &C_{\circ}^c \equiv C_3^c / \left( \langle \hat{h}_3 \rangle \sigma\big(\hat{S}_2^z\big) \sigma\big(\hat{S}_1^z\big) \right)  \,.
    \end{aligned}
\end{equation}
The quantity $C_{\circ}^c$ measures the influence of a hole at site $\vb{r}_3$ on the magnetic correlations between sites $\vb{r}_1$ and $\vb{r}_2$, normalized by the hole density and spin variance in order to offer a more appropriate comparison across a range of dopings. For ease of notation, we will define the two quantities $C^{01}_\circ \equiv C_{\circ}^c(\vb{r} + \hat{\vb{x}}, \vb{r} + \hat{\vb{x}} + \hat{\vb{y}}, \vb{r})$ and $C^{11}_\circ \equiv C_{\circ}^c(\vb{r} + \hat{\vb{x}}, \vb{r} + \hat{\vb{y}}, \vb{r})$ as probing the magnetic correlations on the nearest-neighbor bond and next-nearest-neighbor bond, respectively, closest to the hole. Likewise, we define the nearest-neighbor, next-nearest-neighbor, and next-next-nearest-neighbor spin-spin correlation functions as $C^{01}$, $C^{11}$, and $C^{02}$, respectively. These local correlation functions have recently been measured in cold atom experiments~\cite{koepsell2021} at $U/t \approx 8.9$ for spin-spin correlations and $U/t \approx 7.4$ for their hole-induced counterparts. These experiments were conducted at moderate temperatures ($k_B T \approx 0.5t$) where long-range order is absent. By optimizing our variational states against the Hubbard model to obtain low-energy ansatzes, we aim to evaluate the ability of these states to capture observed low-temperature correlations in doped Mott insulators.

\begin{figure}
    \centering
    \includegraphics[width=0.45\textwidth]{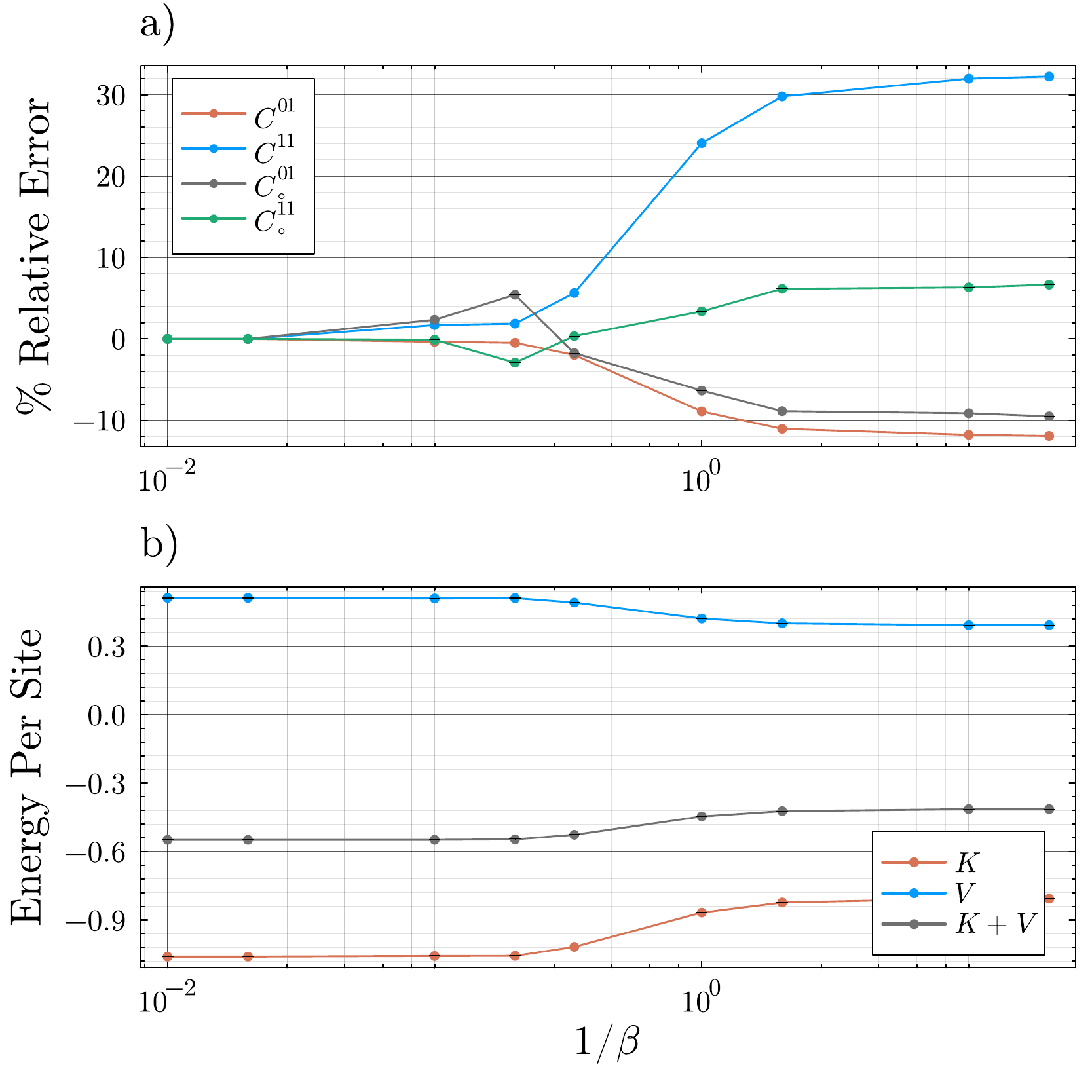}
    \caption{%We plot the magnetic 
    Magnetic correlations (top) and energy (bottom) of our optimized variational states for an $8 \times 8$ lattice at doping $p = 1/16$ for $U/t = 7.4$. These quantities are plotted as a function of the parameter $\beta$ which controls the severity of the sign problem. We find that the energy per site is reduced substantially as $\beta \rightarrow \infty$, driven by a decrease in the average kinetic energy. For the magnetic correlations, both bare and post-selected in the presence of a hole, the relative error as compared to their $\beta = \infty$ value is around 10\%. }
    \label{fig:betaTrend}
\end{figure}

Recall the presence of a sign problem for finite $\beta$, which is a parameter that tunes from our wavefunctions only accounting for a subgroup of the gauge fluctuations ($\beta = 0$) to fully accounting for them ($\beta \rightarrow \infty$). The severity of the sign problem roughly scales with the magnitude of charge fluctuations, and hence increases both at half filling as the charge gap decreases and also %as well as 
with increasing hole doping. For small hole doping, the sign problem remains weak enough ($\langle s \rangle \approx 0.1$) such that the $\beta$-dependence of observables in metallic FL* states can be studied. This is shown in Fig.~\ref{fig:betaTrend} at hole doping $p = 1/16$, where we plot both the energy per site and the nearest-neighbor hole-induced magnetic correlations as a function of $\beta$. We find that, while the variational energy has a strong dependence on $\beta$ and is substantially lowered by performing the full rung singlet projection, both the bare spin-spin correlations and hole-induced spin-spin correlations %only 
display only 
a moderate dependence of around 10\% relative error (with longer-distance spin-spin correlation functions having a larger dependence). We conjecture that this general trend holds true for larger doping, where the sign problem becomes more severe and explicit verification of this trend is not feasible. In analyzing these correlations, we do not focus extensively on energetics beyond using them as a guide to optimizing our wavefunction. This is firstly due to lack of %well-established 
benchmark results at the specific 
%$U / t = 7.4$, 
$U / t$ value, 
but also due to the aforementioned observation that the quantitative value of the variational energy has a strong dependence on $\beta$, and hence we do not expect to obtain competitive energies for small $\beta$. 
%because we find that weakening the sign problem by relaxing the full rung singlet projection incurs a substantial energy penalty. Nevertheless, for half-filling and for very small doping, we are able to verify explicitly that our multi-point correlation functions of interest are much less impacted by this approximation than the variational energy. 

\begin{figure*}
    \centering
    \includegraphics[width=\textwidth]{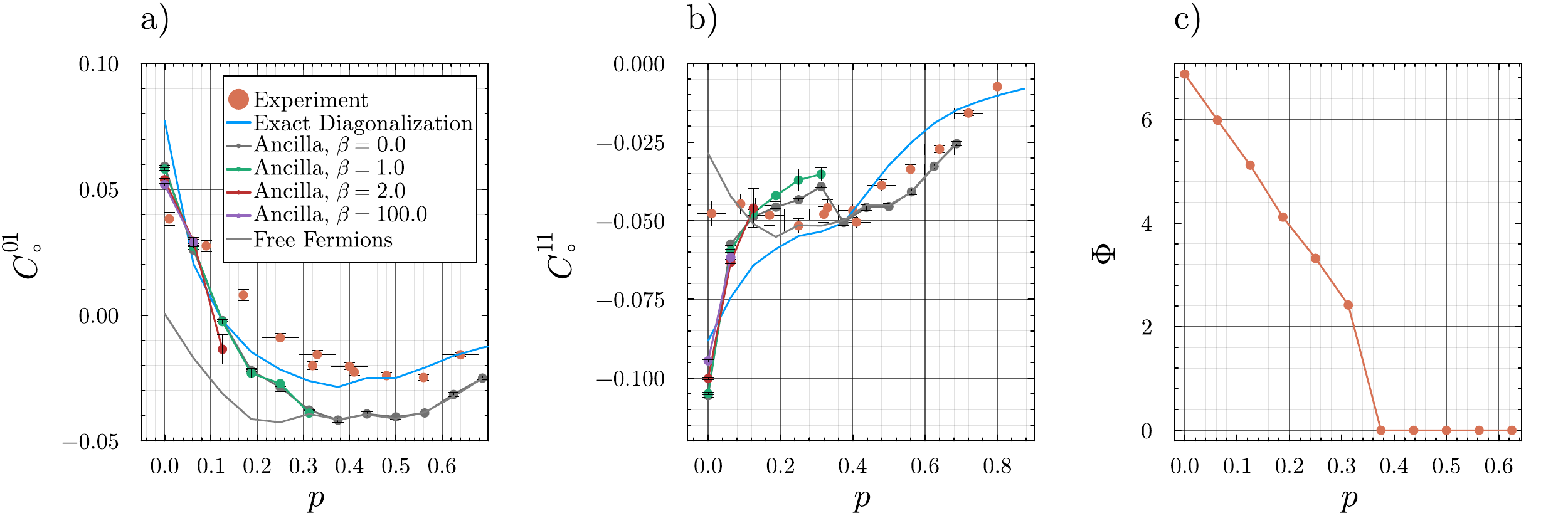}
    \caption{Multi-point correlation functions and 
    comparisons with cold-atom experiment.
    At small hole doping, the presence of a hole induces spin-spin correlations that oppose the antiferromagnetic background order. 
    Panels (a) and  (b) show 
    the nearest neighbor and next-nearest neighbor spin correlations. Quantitative agreement is seen with cold-atom simulations of the Fermi Hubbard model performed in~\cite{koepsell2021}. A crossover to Fermi liquid-like correlations, which become energetically favorable at $p > 0.3125$, also quantitatively captures the behavior of the multi-point correlation functions at high doping.
  %  These multi-point correlation functions for nearest neighbor (a) and next-nearest neighbor (b) spins show quantitative agreement with cold-atom simulations of the Fermi Hubbard model performed in~\cite{koepsell2021}. 
  In (c), we plot the optimized value of the ancilla hybridization $\Phi$ across a range of doping.}
    \label{fig:dopingCorrelations}
\end{figure*}

In Fig.~\ref{fig:dopingCorrelations}, we plot the nearest-neighbor and next-nearest-neighbor spin-spin correlations in the presence of a single hole as a function of hole doping for $\beta = 0\,, 1\,, 2\,, 100$ and $U/t=7.4$. As demonstrated previously in Fig.~\ref{fig:betaTrend}, observables with $\beta = 100$ are essentially indistinguishable from the $\beta \rightarrow \infty$ value. These quantities are obtained by optimizing the energy of our variational FL* wavefunctions against the Hubbard model at each value of doping at $U/t=7.4$. The optimal value of the single variational parameter $\Phi$ is also plotted as a function of doping, where it is seen to decrease linearly from its optimal value of $\Phi \approx U$ at half filling. As the optimal value of $\Phi$ does not change substantially upon increasing $\beta$, we hold fixed the optimal $\beta=0$ value for all $\beta$ to reduce computational costs. While a large sweep of doping is only possible for $\beta = 0$, where the sign problem is absent, we are able to increase $\beta$ for small doping and demonstrate that our correlation functions are not modified substantially by this change.

We find favorable quantitative agreement of our variational ansatzes with experimental data from cold atom experiments collected in~\cite{koepsell2021} as well as exact diagonalization (ED) calculations (also taken from~\cite{koepsell2021})
%simulations 
of the Fermi-Hubbard model at temperatures estimated from the experiment, $k_b T = 0.52t$. % - the ED results are also taken from~\cite{koepsell2021}. 
In particular, our data supports the emergence of magnetic polarons - properties of doped insulators with strong antiferromagnetic fluctuations, where a hole becomes dressed with a cloud of magnetic correlations that opposes the antiferromagnetic background. This is reflected in a positive (ferromagnetic) nearest-neighbor correlation $C^{01}_\circ$, and a negative (antiferromagnetic) next-nearest-neighbor correlation $C^{11}_{\circ}$. Both these values are present at low doping, and our variational wavefunctions also capture the melting into Fermi liquid-like correlations at high doping, where all correlations are antiferromagnetic. We emphasize that these variational wavefunctions are the first to successfully capture this crossover - prior comparisons between cold atom experiments and phenomenological models of doped Mott insulators find that Gutzwiller-projected resonating valence bond and $\pi$-flux states yield only Fermi liquid-like correlations, and ``string'' models of magnetic polarons are unable to capture the Fermi liquid crossover at higher doping~\cite{koepsell2021}.

\begin{figure*}
    \centering
    \includegraphics[width=\textwidth]{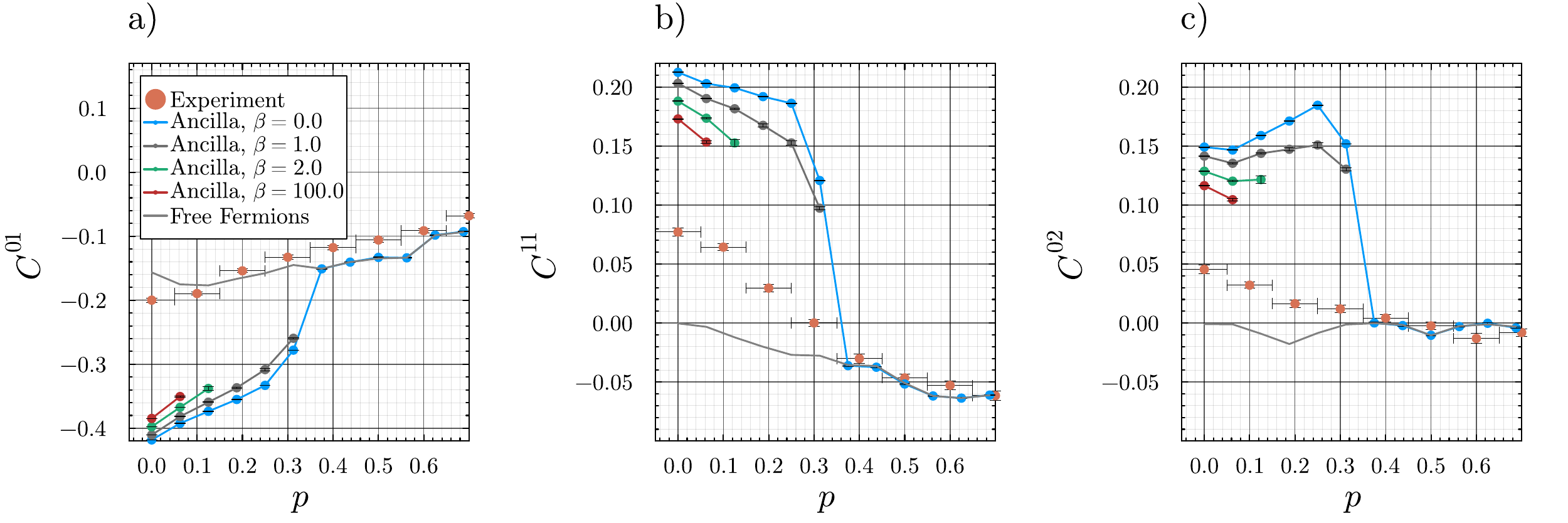}
    \caption{%We plot the nearest-neighbor, next-nearest-neighbor, and next-next-nearest neighbor spin-spin correlation functions of our optimized FL* wavefunctions as a function. 
    Nearest-neighbor, next-nearest-neighbor, and next-next-nearest neighbor spin-spin correlation functions of our optimized FL* wavefunctions.
    All three quantities reflect antiferromagnetic fluctuations which decay as a function of doping, with a sign change in next-nearest-neighbor correlations at $p \approx 0.4$. These trends are consistent with measurements from cold-atom experiment, in particular the crossover from ferromagnetic to antiferromagnetic correlations of next-nearest and next-next-nearest neighbor spins at $p \approx 0.3$.%simulations, 
    Although the quantitative values differ more than for the hole-induced correlation functions in Fig.~\ref{fig:dopingCorrelations}, we note that these observables are more sensitive to our small $\beta$ approximation and do trend towards more agreement with experimental results as $\beta \rightarrow \infty$.}
    \label{fig:spinCorrelations}
\end{figure*}

In Fig.~\ref{fig:spinCorrelations}, we show the bare spin-spin correlation functions for nearest, next-nearest, and next-next-nearest neighbor sites for variational wavefunctions optimized at $U/t = 8.9$. In concurrence with cold atom experiments, the correlations reflect antiferromagnetic order and is gradually suppressed at higher doping. 
The observation from experiment of a
%A 
switch in sign of the next-nearest neighbor correlations around $p \approx 0.3$ is captured in our wavefunctions, albeit at a higher doping. While there is less quantitative agreement with experiment than the hole-induced correlation functions, we find that these quantities depend more strongly on $\beta$, and with increasing $\beta$ trend towards the numerical values predicted by experiment. Given that the experimental measurements are at finite and relatively high temperatures, and our calculations aim for the ground state but with a number of factors 
affecting its accuracy, this level 
of agreement seems reasonable.

\section{Relation to ghost Gutzwiller methods}
Although orthogonal to our main results, we note here that our variational wavefunctions bear close connections to recently-developed ``ghost Gutzwiller'' approximations~\cite{lanata2017} (ghost-GA), where auxiliary orbitals are used to enhance the expressivity of variational wavefunctions and projection back down to the physical Hilbert space is performed using a multi-orbital Gutzwiller approximation, valid in the limit of infinite dimensions. We find that our mean-field wavefunction and rung singlet projection operator satisfy the ``Gutzwiller constraints'' imposed in these ghost-GA studies, and hence are amenable to an analytic Gutzwiller approximation. A study of this approximation is presented in Appendix~\ref{sec:ghost}. We find that these approximations lead to insulating states at all dopings and hence are ineffective at extracting the physics of our wavefunctions away from half filling.
Nevertheless, the accurate description of a Mott insulating state at half-filling leads to insights in the ghost-GA. Namely, it has been previously shown that a single ghost orbital (or ``layer,'' in our terminology) can describe a Mott insulator within the ghost-GA~\cite{lanata2017}. Our three-layer model leads to identical predictions for the electron self-energy; moreover, details of the spin liquid layer drop out in the ghost-GA and one is left with an effective two-band description which parallels prior results. Although a single-band paramagnetic Mott insulator is permitted in infinite dimensions, the Lieb-Schultz-Mattis (LSM) theorem~\cite{lieb1961} and %it's
its
higher-dimensional generalizations~\cite{oshikawa2000, hastings2004} place non-perturbative constraints on ground states of finite-dimensional system. In particular, such an insulating state is disallowed in two dimensions without the coexistence of topological order, which our third spin liquid layer explicitly provides. Hence, our description provides a ``minimal completion'' of the two-band ghost-GA Mott insulator which can be valid for two dimensions down to zero temperature. Such a completion may turn out to be essential if one attempts to compute corrections beyond the limit of infinite dimensions.
\section{Discussion}
The ability to quantitatively assess the properties of fractionalized metallic phases in single-band electron models through variational Monte Carlo, in analogy to Gutzwiller-projected spin liquid states, is an essential step in characterizing these exotic phases of matter. In this work, we provide such an assessment of fractionalized Fermi liquid phases on the square lattice, using recently-developed ancilla models of fractionalization to construct numerically-tractable wavefunctions. Although a full evaluation of these wavefunctions has a sign problem, we find that the sign problem is sufficiently weak at half filling and low hole doping to perform Monte Carlo simulations. Moreover, by introducing an approximation that reduces the sign problem at the cost of suppressing a particular subset of gauge fluctuations, we demonstrate that spin correlations are only weakly affected by this approximation. This allows for a study of these fractionalized phases over a wide range of doping, with which we demonstrate the emergence of magnetic polaronic correlations at low doping which quantitatively agree with recent low-temperature cold-atom simulations of the Fermi-Hubbard model~\cite{koepsell2021}.

Our work has focused primarily on fractionalized metallic phases, where analytical methods are restricted. A related application of these variational wavefunctions is in describing spin liquid states that can arise at half-filling near metal-insulator transitions, such as the chiral spin liquid in the triangular lattice Hubbard model~\cite{szasz2020}. These states possess gapped charge excitations and qualitatively share the same features as analogous spin liquids obtained in the Heisenberg limit; however, capturing these charge degrees of freedom is essential in obtaining competitive variational energies as compared to unbiased studies such as DMRG. Prior attempts to develop variational wavefunctions for these spin liquids using singular Jastrow factors~\cite{tocchio2021} have led to symmetry-broken states and non-competitive energies, which our method would avoid. As these phases have gapped charge excitations, we anticipate the sign problem to be relatively weak in these parameter regimes.  

In this work our goal was to show, as a proof of concept, that a wide class of correlated wave functions produced by the ancilla ansatz can be evaluated and optimized. 
For the future, many technical improvements are possible, both in 
the implementation of the ansatz and the Monte Carlo sampling algorithm. For example, many more variational degrees of freedom can be introduced. The sign problem can be handled in different ways (e.g., by imposing constraints instead of using relatively small $\beta$). 
The sampling could be performed in 
alternative manifolds (e.g., in 
Slater determinant versus occupancy). MPS-based approaches for evlauting ancilla wavefunctions, which were studied in one dimension~\cite{zhou2023}, may support generalizations to higher dimensions, which would provide complimentary results.

We note a related paper~\cite{muller2024} which also analyzes these ancilla wavefunctions in the $\beta = 0$ limit and will appear jointly on arXiv. Our results are consistent with each other.
\section{Acknowledgment}

We thank Subir Sachdev, Yasir Iqbal, Tobias M{\"u}ller, Naoto Nagaosa, and Ryan Levy for helpful discussions. H.S is supported by the U.S. National Science Foundation grant No. DMR-2245246.
The Flatiron Institute is a division of the Simons Foundation.

\appendix
\section{Numerical details}
\label{app:numerics}
Here, we provide additional details on our numerical simulations. As described in the main text, our wavefunction ansatz is defined by a mean-field state $\ket{\psi_{c, f_1}^{\text{MF}}} \otimes \ket{\psi_{f_2}^{\text{MF}}}$ and a projection operator
\begin{equation}
  \begin{aligned}
    \mathcal{P}_\beta &\equiv e^{-\beta \sum_i \left( \vb{S}_{f_1, i} + \vb{S}_{f_2, i} \right)^2 } \mathcal{P}_{S^z = 0}\,,
     \\
     \mathcal{P}_{S^z = 0} &\equiv \sum_{x} \ket{x}\bra{x}\,.
  \end{aligned}
\end{equation}
where $\ket{x}$ enumerates all real-space configurations of ancilla fermions in the $S^z$ basis where each site is singly occupied and $S^z = 0$ on each rung.
Our wavefunctions of interest are given in the $\beta \rightarrow \infty$ limit, and the magnitude of $\beta$ dictates the severity of the sign problem in our simulations.

We enumerate a basis of states in the extended Hilbert space (physical and ancilla fermions) by $\ket{a, x}$, where $a$ denotes a real-space configuration of electrons $c$ and $x$ is a configuration of ancilla fermions in the basis described in the previous paragraph. The expectation value of an observable $\mathcal{O}$ is given by
\begin{widetext}
\begin{equation}
  \begin{aligned}
  \label{eq:observable}
    \langle \mathcal{O} \rangle &= \frac{\bra{\psi^{\text{MF}}} \mathcal{P}_{\beta}^\dagger \mathcal{O} \mathcal{P}_{\beta} \ket{\psi^{\text{MF}}}}{\bra{\psi^{\text{MF}}} \mathcal{P}_{\beta}^\dagger \mathcal{P}_{\beta} \ket{\psi^{\text{MF}}}} 
  =   \frac{\bra{\psi^{\text{MF}}}\left[ \sum_{x, y, a} \ket{a, x}\bra{a, x} e^{-2\beta \sum_i \left( \vb{S}_{f_1, i} + \vb{S}_{f_2, i} \right)^2 } \ket{a, y} \bra{a, y} \right] \mathcal{O}\ket{\psi^{\text{MF}}}}{\bra{\psi^{\text{MF}}}  \left[\sum_{x, y, a} \ket{a, x}\bra{a, x} e^{-2\beta \sum_i \left( \vb{S}_{f_1, i} + \vb{S}_{f_2, i} \right)^2 } \ket{a, y} \bra{a, y} \right]\ket{\psi^{\text{MF}}}} \\
%=   \frac{\bra{\psi^{\text{MF}}}\left[ \sum_{x, y, a,a'} \ket{a, x}\bra{a, x} e^{-\beta \sum_i \left( \vb{S}_{f_1, i} + \vb{S}_{f_2, i} \right)^2 }\,\mathcal{O}\, e^{-\beta \sum_i \left( \vb{S}_{f_1, i} + \vb{S}_{f_2, i} \right)^2 } \ket{a', y} \bra{a', y} \right] \ket{\psi^{\text{MF}}}}{\bra{\psi^{\text{MF}}}  \left[\sum_{x, y, a,a'} \ket{a, x}\bra{a, x} e^{-2\beta \sum_i \left( \vb{S}_{f_1, i} + \vb{S}_{f_2, i} \right)^2 } \ket{a', y} \bra{a', y} \right]\ket{\psi^{\text{MF}}}} \\
  &=  \frac{\bra{\psi^{\text{MF}}}\left[ \sum_{x, y, a} e^{-d(x, y)} \ket{a, x} \bra{a, y} \right] \mathcal{O}\ket{\psi^{\text{MF}}}}{\bra{\psi^{\text{MF}}}  \left[\sum_{x, y, a} e^{-d(x, y)}\ket{a, x}\bra{a, y} \right]\ket{\psi^{\text{MF}}}} 
  %&=  \frac{\bra{\psi^{\text{MF}}}\left[ \sum_{x, y, a,a'} e^{-d(x, y)} \ket{a, x}  \,\bra{a, x}\mathcal{O}\ket{a', y} \,\bra{a', y} \right]
  %\ket{\psi^{\text{MF}}}}{\bra{\psi^{\text{MF}}}  \left[\sum_{x, y, a} e^{-d(x, y)}\ket{a, x}\bra{a, y} \right]\ket{\psi^{\text{MF}}}} 
  = \frac{\sum_{x, y, a} p(x, y, a) s(x, y, a) \mathcal{O}(x, y, a)}{\sum_{x, y, a} p(x, y, a) s(x, y, a)}\,, \\
  \end{aligned}
\end{equation}
\end{widetext}
where we define
\begin{equation}
  \begin{aligned}
    W(x, y, a) &= \braket{\psi^{\text{MF}}}{a, x}\braket{a, y}{\psi^{\text{MF}}} \,, \\
    p(x, y, a) &= e^{-d(x, y)}\abs{W(x, y, a)} \,,\\
 %   s(x, y, a) &= \arg\left[W(x, y, a)\right]\,, \\
    s(x, y, a) &= \frac{W(x, y, a)}{\abs{W(x, y, a)}}\,, \\
    \mathcal{O}(x, y, a) &= \frac{\bra{a, y} \mathcal{O} \ket{\psi^{\text{MF}}}}{\braket{a, y}{\psi^{\text{MF}}}}\,. \\
   % \mathcal{O}(x, y, a) &= \sum_{a'} \bra{a, x} \mathcal{O}\ket{a',y}\,\frac{\braket{a', y}{\psi^{\text{MF}}}}{\braket{a, y}{\psi^{\text{MF}}}}\,. \\
  \end{aligned}
\end{equation}
For any operator $\mathcal{O}$ diagonal in $\{a\}$ 
(configuration basis), $\mathcal{O}(x, y, a) $ in the last equation reduces to 
$\bra{a, y} \mathcal{O}\ket{a,y}$.
Note that $\mathcal{O}(x, y, a)$ is actually independent of $x$; however, since $p(x, y, a) = p(y, x, a)$, one can symmetrize all quantities with respect to $x \leftrightarrow y$. The distance function $d(x, y)$ in Eq.~\ref{eq:observable} controls the severity of the sign problem, and is given by
\begin{equation}
  \begin{aligned}
    e^{-d(x, y)} &\equiv \bra{a, x} e^{-2\beta \sum_i \left( \vb{S}_{f_1, i} + \vb{S}_{f_2, i} \right)^2 } \ket{a, y} \\
    &=  2^{-N} \left( 1 + e^{-2\beta} \right)^{N - \tilde{d}(x, y)} (1 - e^{-2\beta})^{\tilde{d}(x, y)}
  \end{aligned}
\end{equation}
where $\tilde{d}(x, y)$ is the Manhattan distance between the two configurations and counts the number of rungs in $x$ that differ from $y$. For $\beta = 0$, $e^{-d(x, y)} = \delta_{x, y}$ and the average sign is $1$. For $\beta \rightarrow \infty$, $e^{-d(x, y)} \propto 1 $ and the sign is uncontrolled. In practice, we find the sign problem to be relatively weak at half filling and for small hole doping, suggesting that the severity of the sign problem is correlated with the strength of charge fluctuations.

A state in our Monte Carlo algorithm is specified by a configuration of physical electrons, $a$, as well as two configurations of the ancilla spins $x$, $y$, corresponding to the left and right set of states inserted in Eq.~\ref{eq:observable}. We run the Metropolis algorithm with four types of updates:
\begin{itemize}
    \item An electron hopping in $a$
    \item A spin exchange in $a$
    \item A spin exchange on either of the ancilla configurations $x$, $y$
    \item A simultaneous spin exchange on $x$ and $y$
    \item A simultaneous spin exchange on $x$, $y$, and $a$. 
\end{itemize}
In the Heisenberg limit at half-filling, only the final move has a non-zero probability and we essentially recover a standard variational Monte Carlo simulation of a Gutzwiller-projected spinon state.

Optimization of the variational parameter $\Phi$ is performed using gradient descent. The presence of a sign problem non-trivially modifies the expression for the energy derivative, as demonstrated in~\cite{sorella2022} - this modified expression is used in our optimization at half-filling.

There is a subtlety in sampling statistics which leads to an increased error in regimes, related to the infinite variance problem discussed in~\cite{shi2016}. Recall that our probability density $\abs{\bra{\psi^{\text{MF}}} \ket{a,x} \bra{a,y} \ket{\psi^{\text{MF}}}}$ depends on both the overlap of the mean-field wavefunction with the configurations $\ket{a,x}\,, \ket{a,y}$. However, in our construction, the local operator $\mathcal{O}(a, x, y) = \frac{\bra{a, y} \mathcal{O} \ket{\psi^{\text{MF}}}}{\braket{a, y}{\psi^{\text{MF}}}}$ only depends on $\bra{a, y}$. It is possible for the denominator of this expression to be quite small, leading to an anomalously large local operator. Crucially, because the probability density depends \textit{linearly} on the denominator - rather than quadratically for a more standard VMC when $x = y$ - such anomalous samples are only weakly suppressed and require longer sampling in order to resolve. This issue is observable-dependent, as the numerator of $\mathcal{O}(a, x, y)$ may be sufficiently correlated with the denominator being small such that this is not an issue - this is the case if $\mathcal{O}$ is diagonal in the basis $\ket{a, y}$. 
\section{Details of the ghost Gutzwiller calculation}
\label{sec:ghost}
Ghost-Gutzwiller (ghost-GA) approximations are an increasingly common tool in studying properties of strongly-correlated systems in a numerically-efficient manner~\cite{lanata2017, frank2021, lanata2022, lee2023c, lee2023d, lanata2023, mejuto-zaera2023}. In these methods, auxiliary ``ghost'' orbtials are introduced in a mean-field ansatz, and a projection operator is applied to generate a correlated wavefunction. This projection is performed using a generalized Gutzwiller approximation. Our FL* wavefunctions bear a strong resemblance to this method, and moreover our projections are amenable to the ghost-GA, which we elaborate on here.

For our FL* wavefunctions at half-filling, we have as input the local particle occupations,
\begin{equation}
\begin{aligned}
    \bra{\psi_0} c_{i\sigma}^\dagger c_{i\sigma}^\pdagger \ket{\psi_0} &= 0.5
    \\
    \bra{\psi_0} f_{1i\sigma}^\dagger f_{1i\sigma}^\pdagger \ket{\psi_0} &= 0.5
    \\
     \bra{\psi_0} f_{2i\sigma}^\dagger f_{2i\sigma}^\pdagger \ket{\psi_0} &= 0.5
    \\
    \bra{\psi_0} c_{i\sigma}^\dagger f_{1i\sigma}^\pdagger \ket{\psi_0} &= \alpha
    \label{eq:particleOccs}
    \end{aligned}
\end{equation}
with all other occupations set to zero. The tuneable parameter $\alpha$ controls the degree of hybridization between the physical and ancilla fermions. In our particular choice of tight-binding model in the main text, each filled single-particle eigenstate is an equal-weight superposition of $c$ and $f_1$ fermions with no relative phase factor, so a value $\alpha = 0.5$ is actually fixed for any non-zero hybridization strength $\Phi > 0$. We will admit a variable $\alpha$ as this can in principle be accessed by more sophisticated tight binding models, and moreover because the self-consistent determination of our mean-field Hamiltonian from a Hubbard model~\cite{mascot2022} identifies $\langle c_{i\sigma}^\dagger f_{1i\sigma}^\pdagger \rangle \sim \Phi$.

These local occupations can be diagonalized in the ``natural'' basis. At this point, we choose the variables $c_{i\alpha}$ to refer to all fermions, both physical and ancilla, in the original basis (with $\alpha=1\,,6$ labelling both spin and type of fermion), and $d_{ia}$ as fermions in the natural basis, such that
\begin{equation}
     d_{ia}^\dagger \equiv \sum_\alpha \left[U_i\right]_{\alpha a} c_{i\alpha}^\dagger 
\end{equation}
and
\begin{equation}
    \bra{\psi_0} d_{ia}^\dagger d_{ib}^\pdagger \ket{\psi_0} \equiv \delta_{ab} \left[ n_i^0 \right]_{bb}\,.
\end{equation}
Our rung-singlet projection is parameterized by the matrix $\Lambda_{\Gamma n}$ such that
\begin{equation}
    \mathcal{P}_i = \sum_{\Gamma n} \Lambda_{\Gamma n} \ket{\Gamma, i} \bra{n, i}
\end{equation}
where $\ket{\Gamma, i}$ and $\ket{n, i}$ are Fock states in the original and natural basis, respectively. Upon defining the following matrices in the local many-body Fock space
\begin{equation}
    \begin{aligned}
        \left[P^0_i\right]_{n n'} &\equiv \braket{\psi_0}{n', i} \braket{n, i}{\psi_0}\,,
        \\
        [D_{ib}]_{nn'} &\equiv \bra{n, i} d_{ib} \ket{n', i}\,,
        \\
         [C_{i\beta}]_{\Gamma \Gamma'} &\equiv \bra{\Gamma, i} c_{i\beta} \ket{\Gamma', i}\,,
         \\
         \phi_i &\equiv \Lambda_i \sqrt{P_i^0}\,,
    \end{aligned}
\end{equation}
we can express our Gutzwiller constraints as
\begin{equation}
    \begin{aligned}
      \Tr\left[ \phi_i^\dagger \phi_i^\pdagger \right] &= 1\,,
      \\
      \Tr\left[ \phi_i^\dagger \phi_i^\pdagger D_{ia}^\dagger D_{ib}^\pdagger \right] &= \delta_{ab} [n_i^0]_{bb}\,.
    \end{aligned}   
\end{equation}
One can verify by explicit numerical evaluation that these constraints are satisfied for our choice of $\ket{\psi_0}$ and $\mathcal{P}_i$, although the first only serves to normalize our projection operator. From this, the renormalization factors $[\mathcal{R}_i]_{a\alpha}$ can be expressed as
\begin{equation}
    [\mathcal{R}_i]_{a \alpha} \equiv \Tr [ P_i^0 \Lambda_i^\dagger D_{i\alpha}^\dagger \Lambda_i^\pdagger D_{ia}^\pdagger] / [n_i^0 ]_{aa}\,.
\end{equation}
A chemical potential $\lambda_{ab}$ in the natural basis is also introduced to self-consistently ensure that the ground state of the quasiparticle Hamiltonian in the natural basis,
\begin{equation}
    H_{\text{qp}} = \sum_k \sum_{ab} \left[ \mathcal{R} \epsilon_k \mathcal{R}^\dagger + \lambda \right] f_{ka}^\dagger f_{kb} 
\end{equation}
satisfies the local particle occupations given in Eq.~\ref{eq:particleOccs}. The energy dispersion $\epsilon_k$ is taken from our physical model of interest, which is a square lattice with nearest-neighbor hopping.

From this, we can obtain the energy-resolved Green's function
\begin{equation}
    G(\epsilon_k, \omega)_{\alpha \beta} = \left[ \mathcal{R}^\dagger \frac{1}{\omega - \mathcal{R} \epsilon_k \mathcal{R}^\dagger + \lambda} \mathcal{R}\right]_{\alpha \beta}
    \,, \quad \alpha\,, \beta = 1,2
\end{equation}
A non-trivial aspect of this matrix equation is that the matrices $\mathcal{R}\,, \lambda$ are in the enlarged Hilbert space and hence $6 \times 6$ matrices, whereas the physical degrees of freedom are restricted to the first two indices (spin up and spin down of the physical electrons). This generates non-linearities in the electron self-energy.

\begin{figure*}
    \centering
    \includegraphics[width=0.45\linewidth]{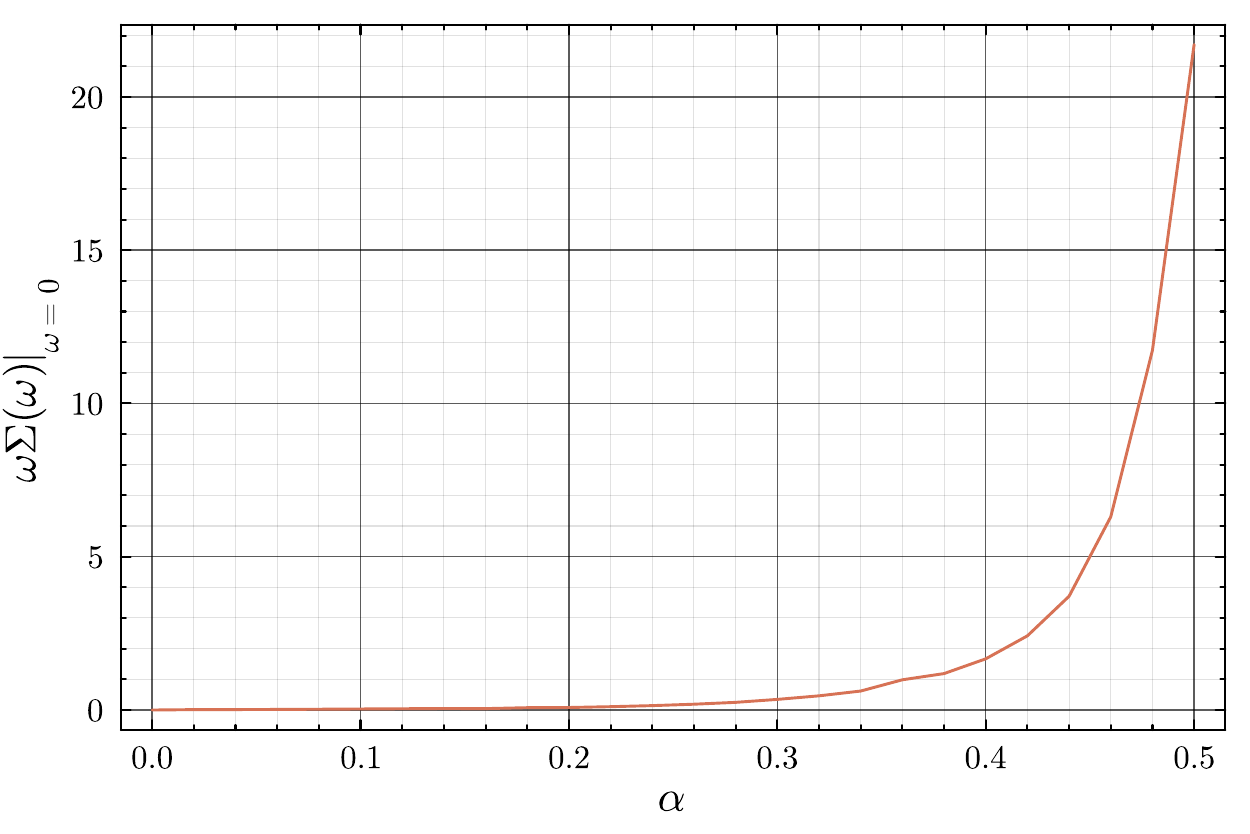}
    \includegraphics[width=0.45\linewidth]{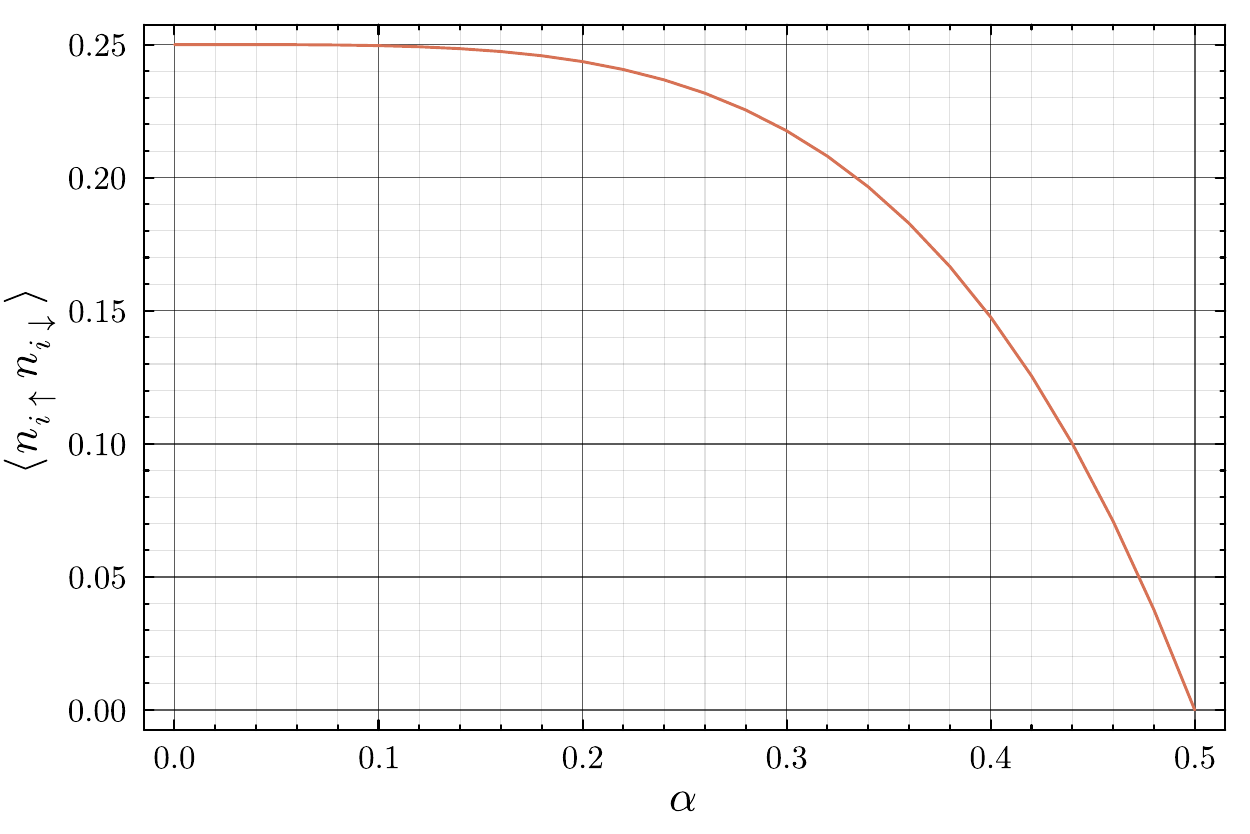}
    \caption{Within the ghost Gutzwiller approximation, we plot the magnitude of the $1/\omega$ pole in the electron self-energy (left) and the average double occupancy (right) of the Mott inuslating states captured by our ancilla model at half filling, as a function of a tunable parameter $\alpha$.}
    \label{fig:ghost}
\end{figure*}
At half-filling, we find that the third spin liquid layer effectively drops out completely from the renormalization matrices, i.e. $\left[\mathcal{R}_i \right]_{a \alpha}$ when $\alpha$ resides in the third layer. In other owords, we effectively reduce to a two-band description of this state. Evaluation of this electron Green's function gives a Mott insulator with a diagonal self-energy of the form 
\begin{equation}
    \Sigma(\omega) = - t\left[a \omega +  \frac{b}{\omega}\right]
\end{equation}
with constants $a$ and $b$, where we have factored out the overall energy scale set by the nearest-neighbor hopping $t$. In our calculations, we find $a = \sqrt{2}$ and $b$ a function of $\alpha$, plotted in Fig.~\ref{fig:ghost} in addition to the average double occupancy. The frequency dependence of this self-energy is identical to the self-energy found in a two-band description of a Mott insulator (i.e. one ghost orbital) in~\cite{lanata2017}. Despite our model having one more layer than~\cite{lanata2017}, we find that the spin liquid layer effectively drops out of our calculations, in that the renormalization matrices $\mathcal{R}$ contain no entries in the spin liquid layer. As a result, general properties of these insulators match the two-band description described in~\cite{lanata2017}. As discussed in the main text, one may think of our wavefunctions as a ``minimal extension'' of the two-band model which is able to satisfy the LSM theorem in low dimensions.  

Away from half-filling, one can still perform a ghost-GA. The electron or hole doping is accounted for in modifying the local occupations in Eq.~\ref{eq:particleOccs}. However, running the ghost-GA results in effective two-band insulating states for all doping. This is driven by the fact that ghost-GA is only able to generate a momentum-independent self-energy and is therefore unable to capture the small Fermi surface present in the mean-field evaluation of the electron Green's function.

\bibliography{ancilla.bib}
\end{document}